\documentclass[10pt]{iopart}
\usepackage{epsfig,latexsym}
\usepackage{graphicx}
\usepackage{dcolumn}
\usepackage{bm}
\newcommand{\be}{\begin{eqnarray}}
\newcommand{\ee}{\end{eqnarray}}

\begin{document}

\title{Quark Mass and the QCD Transition}

\author{\'{A}gnes M\'{o}csy \footnote[1]{Email:
(mocsy@th.physik.uni-frankfurt.de)}}

\address{Frankfurt Institute for Advanced Studies,
J.W.Goethe-Universit\"at\\  Postfach 11 19 32, 60054 Frankfurt am
Main, Germany}

\begin{abstract}
We discuss different aspects of the quark mass dependence of the
chiral and deconfinement transitions. We make predictions for
locating the deconfining critical point, by analyzing the $m_q$
dependence of the ratio of imaginary and real mass of the Polyakov
loop. We then present an effective theory that allows the
extraction of coupling constants from available lattice data for
the temperature dependence of the chiral and Polyakov loop
condensates at different quark masses.
\end{abstract}



Theoretical developments accompanied by recent numerical results
from the lattice have advanced our understanding of QCD. There are
still numerous important questions however that need to be
answered. For instance: Where does the QCD point lie in the
($m_s$,$m_q$) and in the ($T$,$m_q$) phase diagrams? Which
mechanism drives the transition for a given quark mass? Is the QCD
transition more deconfining or chiral symmetry restoring? See
\cite{karsch} for related discussion. As a step towards answering
these questions, first, we discuss the location of the deconfining
critical point in the ($T$,$m_q$) plane. Then, we introduce a
general theory for the Polyakov loop and chiral field at finite
quark mass, and discuss its fit to lattice data.

The pure gauge sector of QCD with $N$ colors, i.e. $m_q=\infty$,
has global $Z_N$ symmetry associated with the center of the
$SU(N)$ gauge group. The Polyakov loop $\ell$, charged under
$Z_N$, serves as order parameter for the deconfining transition.
$\ell$ is real for $N=2$ and the $Z_2$ symmetry breaking
deconfinement transition is of second order. For $N=3$ the
Polyakov loop is complex and transforms as $\ell\rightarrow
\mbox{exp}(2\pi i/3)\ell$. Accordingly, the expectation value
$\langle\ell\rangle=0$ at low temperatures, when $Z_3$ is
unbroken, and $\langle\ell\rangle\neq 0$ above the deconfinement
critical temperature $T_d~$. The deconfinement transition for
$N=3$ is weakly first order. This has been verified by lattice QCD
\cite{karsch}.

In \cite{Pisarski:2000eq} an effective theory for the Polyakov
loops, the Polyakov Loop Model (PLM) has been
constructed\footnote{Some other approaches to study deconfinement
are in \cite{Kogan:2002yr}.}. The potential for $N=3$, up to
quartic terms, is
\be
V(\ell)=-b_1\frac{\ell+\ell^*}{2}-\frac{b_2}{2}|\ell|^2
-\frac{b_3}{3}\frac{\ell^3+\ell^{*3}}{2} +
\frac{1}{4}(|\ell|^2)^2\, .   \label{pot} \ee In a mean field
analysis all coupling constants, except $b_2(T)$, are temperature
independent. There are two masses, associated with the real
$\ell_r=\mbox{Re}~\ell$ and imaginary $\ell_i=\mbox{Im}~\ell$
parts of the Polyakov loop. These are defined from the two-point
functions
\be
\!\!\!\!\!\!\!\!\!\langle\ell_r(x)\ell_r(0)\rangle - \ell_0^2 \sim
\frac{\exp(-m_r x)}{x}\, , \qquad \langle\ell_i(x)\ell_i(0)\rangle
\sim \frac{\exp(-m_i x)}{x}\, , \ee and thus characterize the
exponential tail of the correlators. $\langle\ell\rangle=\ell_0>0$
is real. Denote the two minima $\ell^-=\ell(T_d^-)$ and
$\ell^+=\ell(T_d^+)$. The masses at $T_d^+$ are then
\be
m_r^2 = - b_2 - 2b_3\ell^+ + 3(\ell^+)^2\, ,\qquad m_i^2 = - b_2 -
2b_3\ell^+ + (\ell^+)^2\, . \ee

For $b_1=0$ the potential is $Z_3$ symmetric. The solutions are
\be
\ell_0^- = 0\, , \qquad \ell_0^+ = \frac{2}{3}b_3\, , \qquad b_{2}
= -\frac{2}{9}b_3^2  \, , \label{init}\ee and the ratio of the
masses at $T_d^+$ is $m_i/m_r=3$ \cite{Dumitru:2002cf}. This value
is twice the one expected from lowest order perturbative
calculations, valid at high temperatures. The PLM thus predicts an
increase of $m_i/m_r$ as $T\rightarrow T_d^+$, prediction
supported by the lattice \cite{Datta:2002je}. In principle,
$m_i/m_r$ can receive contributions from pentic and hexatic terms
(see discussion in \cite{Dumitru:2002cf}).

\begin{figure}[htbp]
\begin{center}
\begin{minipage}[t]{5cm}
\epsfig{file=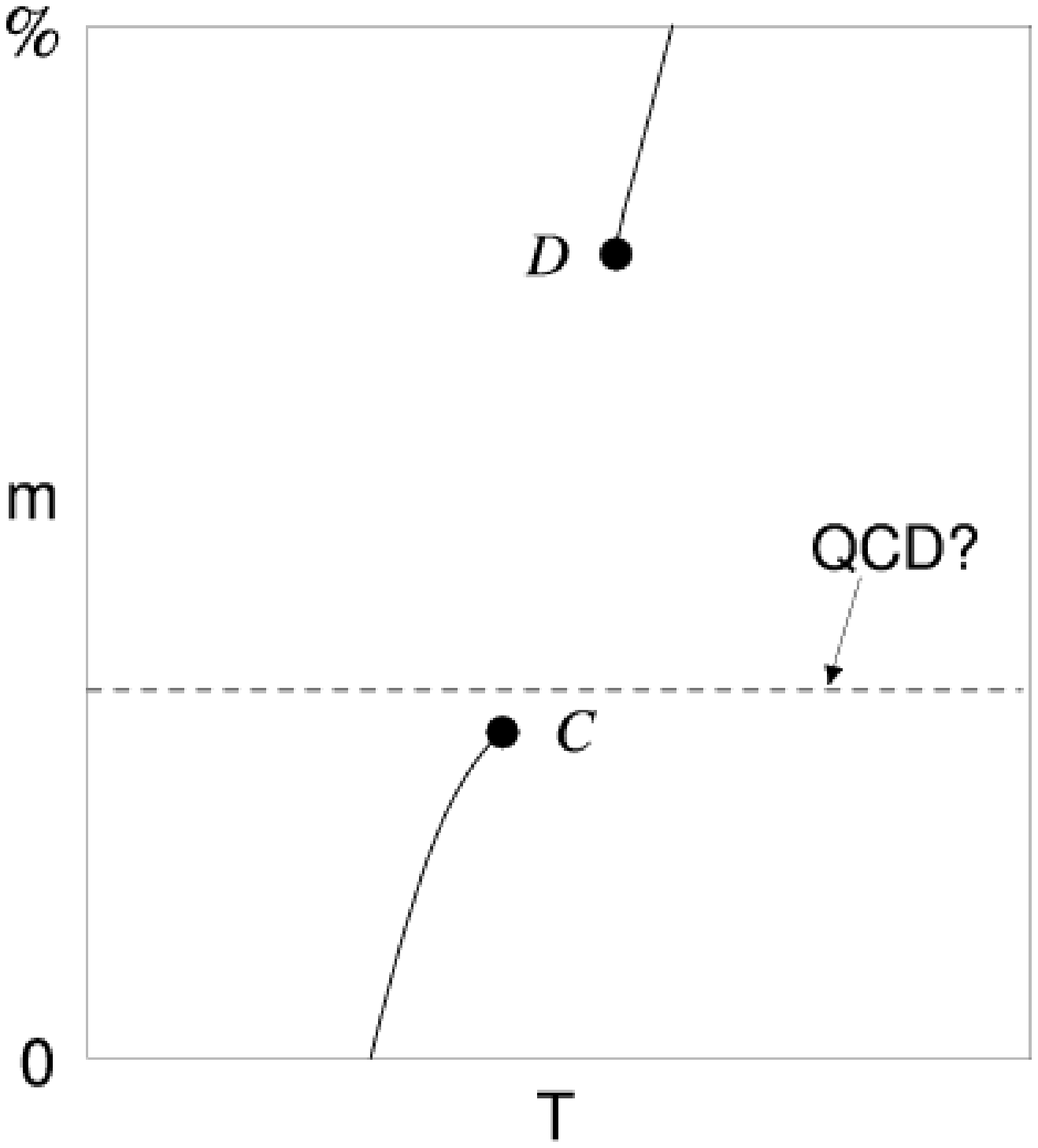,height=50mm}
\end{minipage}
\begin{minipage}[t]{5cm}
\epsfig{file=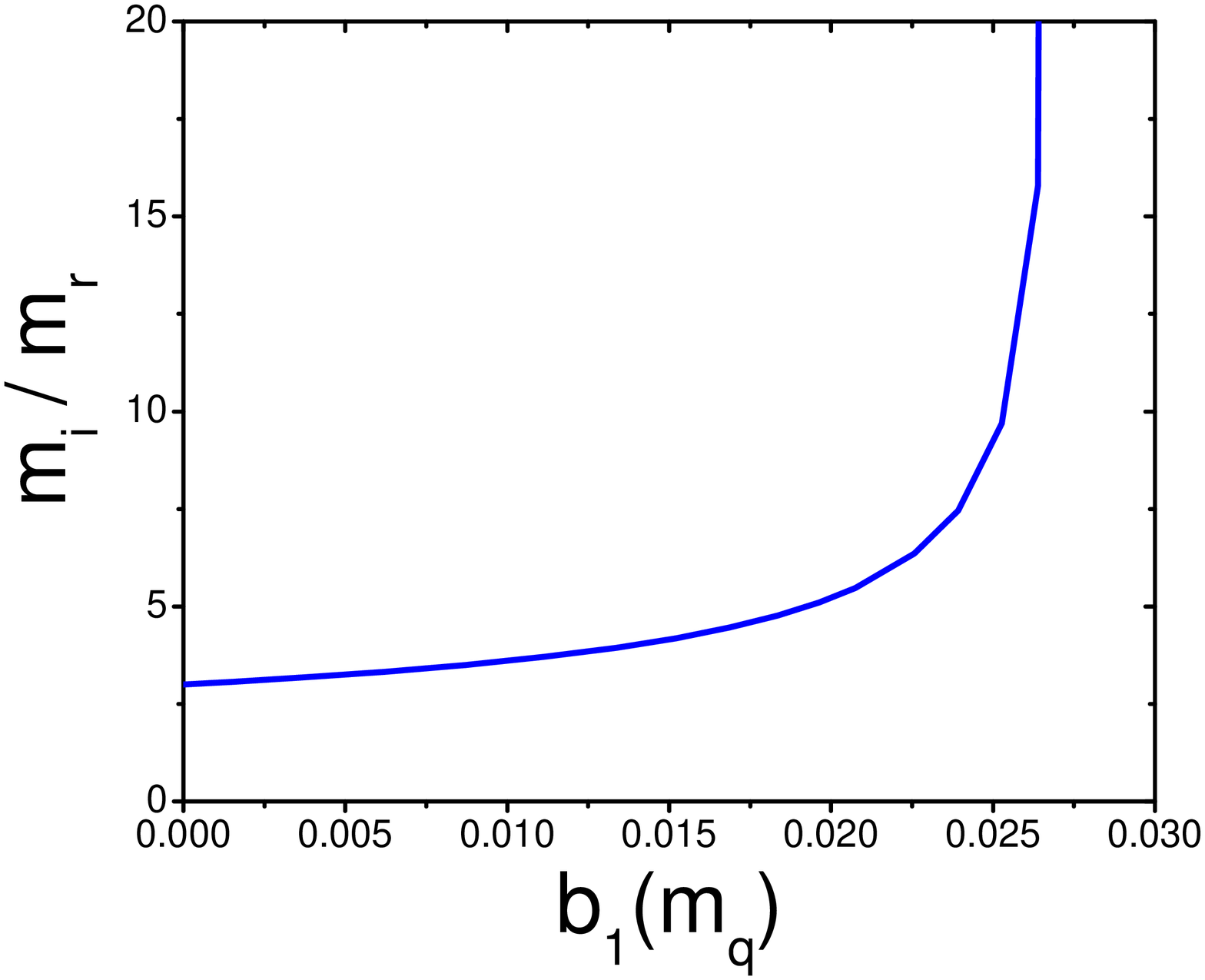,height=45mm}
\end{minipage}
\caption{Left panel: Temperature--quark mass phase diagram. From
\cite{Gavin:1993yk}. Right panel: Quark mass dependence of the
ratio of imaginary and real mass of the Polyakov loop at the
corresponding critical temperature. } \label{fig}
\end{center}
\end{figure}

What happens to $m_i/m_r$ when dynamical quarks are added in the
theory? How does this ratio change for finite quark masses? For
any finite $m_q$ the $Z_3$ symmetry is explicitly broken, and the
linear term $b_1$ is turned on in (\ref{pot}). This corresponds to
tilting the potential and shifting the degenerate minima at $T_d$
towards each other, and thus the transition becomes weaker first
order. An increasing symmetry breaking is equivalent to decreasing
the quark masses from infinity. An explicit functional relation
for $b_1(m_q)$ fitted for lattice data, together with the $m_q$
dependence of $T_d$ is discussed in \cite{Dumitru:2003cf}. For
some value of $m_q$ there is only one minima and the transition is
of second order. This is the deconfining critical point D, shown
in the left panel of Fig. \ref{fig}. When further decreasing the
quark mass the phase transition becomes a crossover. To determine
the influence of quark mass we derived a set of differential
equations
\be
\!\!\!\!\!\!\!\!\!\!\!\!\!\!\!\!\!\!\!\!\!\!\!\!\!\!\!
\!\!\!\!\!\!\!\!\!\!\!\frac{\partial\ell_0^-}{\partial b_1} =
\frac{1}{m_-^2}\left(1+\frac{\partial b_2}{\partial b_1}
\ell_0^-\right)\, ,~~~ \frac{\partial\ell_0^+}{\partial b_1} =
-\frac{1}{m_+^2}\left(1+\frac{\partial b_2}{\partial b_1}
\ell_0^+\right)\, ,~~~\frac{\partial b_2}{\partial b_1} =
-\frac{2}{\ell_0^- + \ell_0^+} \label{b2}\ee and solve them
together with the initial conditions (\ref{init}) numerically.
Here $V''(\ell^-)= m_-^2$ and $V''(\ell^+)=m_+^2$, and $b_3=0.9$.
Note that the third term is always negative, leading to decreasing
$T_d$ with $m_q$ (see the slope of the upper 1st order critical
line in the left panel of Fig.~\ref{fig}). The mass ratio at the
critical temperature in terms of the symmetry breaking is shown in
the right panel of Fig.~\ref{fig}. We find that as the quark mass
decreases from infinity the ratio slightly increases. The second
order deconfining critical point D is reached for $b_1=0.027$ and
here the mass ratio becomes divergent. This means that only the
real part $\ell_r~$, not the $\ell_i~$, becomes massless at D, and
thus $Z_3$ is a symmetry group here. Following the behavior of
$m_i/m_r$ on the lattice would allow for locating D, and also to
determine the quark mass that this point corresponds to.

Another well studied sector of QCD is in the limit of zero and
small quark masses. In this regime $Z_N$ is always broken, but
chiral symmetry is restored above a critical temperature $T_c$.
The order parameter is the chiral condensate $\sigma$. The
transition is second order for two, and first order for three
flavors of massless quarks. For finite quark masses chiral
symmetry is explicitly broken. Increasing $m_q$ from zero means
weakening the first order phase transition. For some $m_q$ the
first order critical line turns into a second order chiral
critical point, C in the left panel of Fig.~\ref{fig}. For even
larger $m_q$ the transition is a crossover. Effective field field
theories have been used to analyze this sector
\cite{Scavenius:2000qd}.

Deconfinement and chiral symmetry restoration are two phenomena
very different in nature. For realistic quark masses both the
chiral and $Z_N$ symmetries are explicitly broken, and likely that
both transitions are in the crossover domain \cite{Gavin:1993yk}.
Here neither $\sigma$, nor $\ell$ can play the role of a true
order parameter. However, lattice simulations show that the
susceptibilities associated with these two quantities peak at the
same temperature, when quarks are in the fundamental
representation of the gauge group \cite{Karsch:1998qj}, indicating
that the transitions occur at the same critical temperature.
Similar results were obtained also in terms of the chemical
potential \cite{Alles:2002st}.

In \cite{Mocsy:2003qw} we provided an explanation within a
generalized Ginzburg-Landau theory to why for $m_q=0$ chiral
symmetry restoration leads to deconfinement. The analysis was
based on the following general idea: The behavior of an order
parameter induces a change in the behavior of non-order parameters
at the transition \cite{Mocsy:2003tr}, via the presence of a
possible coupling between the fields, $g_1\ell\sigma^2$. In
\cite{Mocsy:2003qw} we assumed $g_1>0$, which is confirmed by
extracting it from the lattice \cite{FM}.

Here we extend the analysis of \cite{Mocsy:2003qw} for finite
quark masses. We can then study not only the qualitative relation
between deconfinement and chiral symmetry restoration in terms of
quark masses, but also can provide quantitative statements after
extracting the coupling constants from lattice data \cite{FM}.
Recent lattice results for the temperature dependence of the
chiral condensate for three degenerate massive quarks were
reported in \cite{Bernard:2003dk}. The first results for the
behavior of the renormalized Polyakov loop
\cite{Dumitru:2003hp,{Petreczky:2004pz}} as a function of
temperature, under the same lattice conditions, were reported in
\cite{Petreczky:2004pz}.

We now write down the most general renormalizable effective
Lagrangian that can be built from the chiral field $\sigma$ and
the Polyakov loop $\ell$. The contribution to the potential
$V=V_{Pl}+V_{ch}+V_{int}$ from the Polyakov loop is
\be
V_{Pl}(\ell) = g_0\ell - \frac{b_2}{2}\ell^2 - \frac{b_3}{3}\ell^3
+ \frac{1}{4}\ell^4 \, . \ee Here $g_0=-b_1$ \footnote{This
potential is written only for the real part of the Polyakov loop,
since we are interested in the behavior of its expectation value,
which can always be chosen real at $\mu=0$.}. For the chiral
potential we adopt the linear $\sigma$ model, discarding all
fields but the $\sigma$, since in relation to our discussion these
will not play a role \cite{FM}:
\be
V_{ch}(\sigma) = \frac{m^2}{2}\sigma^2 + \frac{\lambda}{4}\sigma^4
- H\sigma\, . \ee
Here $H=f_\pi m_\pi^2~$, with $m_\pi$ and $f_\pi$ the vacuum pion
mass and decay constant. The possible interactions terms are
\be
V_{int}(\ell,\sigma) = g_1\ell\sigma^2 + g_2\ell^2\sigma^2 +
\bar{g}_1\ell^2\sigma + \bar{g}_2\ell\sigma\, . \ee In a mean
field analysis the field expectation values to first order in
$m_q$ are
\be
\langle\ell\rangle &\simeq& - \frac{g_0}{m_\ell^2} -
\frac{g_1}{m_\ell^2}\langle\sigma\rangle^2 -
\frac{\bar{g}_2}{m_\ell^2}H\langle\sigma\rangle \label{eom-l} \ee
and $\langle\sigma\rangle$ is the solution of
\be
\langle\sigma\rangle^3 + \frac{g_1\bar{g}_2}{2\lambda
m_\ell^2}H\langle\sigma\rangle^2 -
\frac{m_\sigma^2}{2\lambda}\langle\sigma\rangle +
\frac{H}{2\lambda}\left(1+ \frac{g_0\bar{g}_2}{m_\ell^2}\right)
&\simeq& 0 \, . \label{eom-sigma} \ee
For $H=0$ we recover the results of \cite{Mocsy:2003qw}, as
expected. For light quarks of three degenerate flavors, the
Polyakov loop mass is identified as the inverse of the screening
radius and is extracted from the lattice \cite{Petreczky:2004pz}.
For the $\sigma$ mass we assume $m_\sigma^2 (T)=a\left(b
T/T_c-1\right)$ as is customary in a Ginzburg-Landau theory. Then
using (\ref{eom-l}) and (\ref{eom-sigma}) we fit the data from
\cite{Petreczky:2004pz} and \cite{Bernard:2003dk}. The couplings
extracted this way confirm the assumption made in
\cite{Mocsy:2003qw} in that $g_1>0$ and $g_0<0$. Accordingly, for
massless quarks ($H=0$) the decrease of the order parameter, the
chiral condensate induces an increase in the Polyakov loop, as one
can see directly in (\ref{eom-l}). Which field drives the
transition in the case of massive quarks can be infered from the
knowledge of the couplings. These results will be presented
elsewhere \cite{FM}.

With our effective model, we will be able to make predictions
relevant for the phenomenology of heavy ion collisions. The
extension of this analysis to the chemical potential axis of the
phase diagram is relevant not only in the context of the critical
point in the $(T,\mu)$ plane, but also in the understanding of the
phase structure of QCD.

\section*{Acknowledgements}

Special thanks to my collaborators A.Dumitru, E.S.Fraga,
F.Sannino, K.Tuominen. I thank P.Petreczky and R.Pisarski for
discussions and A.Dumitru for carefully reading the manuscript. I
am grateful to the Theoretical Physics Institute at J.W.~Goethe
University in Frankfurt for the kind hospitality, where I am a
Humboldt Fellow.


\end{document}